\documentclass[preprint,12pt]{elsarticle}

\usepackage{amssymb}
\usepackage{amsmath}
\usepackage{longtable}
\usepackage{graphicx}
\usepackage{multirow}
\usepackage{rotating}
\usepackage{makecell}
\usepackage{booktabs}
\usepackage{paralist}
\usepackage{microtype}
\usepackage{url}
\usepackage{hyperref}

\journal{}

\begin{document}

\begin{frontmatter}



\title{Blockchain Data Analytics: Review and Challenges}


\author{Rischan Mafrur} 
\ead{rischan.mafrur@mq.edu.au}

\affiliation{organization={Department of Applied Finance, Macquarie University},
            country={Australia}}

\small
\begin{abstract}
The integration of blockchain technology with data analytics is essential for extracting insights in the cryptocurrency space. Although academic literature on blockchain data analytics is limited, various industry solutions have emerged to address these needs. This paper provides a comprehensive literature review, drawing from both academic research and industry applications. We classify blockchain analytics tools into categories such as block explorers, on-chain data providers, research platforms, and crypto market data providers. Additionally, we discuss the challenges associated with blockchain data analytics, including data accessibility, scalability, accuracy, and interoperability. Our findings emphasize the importance of bridging academic research and industry innovations to advance blockchain data analytics.

\end{abstract}




\end{frontmatter}




\section{Introduction}
Blockchain technology, recognized for its decentralized, transparent, and immutable nature, has transformed various sectors, including finance \cite{michael2018blockchain}, supply chain management \cite{queiroz2020blockchain}, healthcare \cite{attaran2022blockchain}, and others \cite{yaga2019blockchain}. However, as blockchain networks expand, the increasing volume, velocity, and variety of generated data pose significant challenges for effective analysis. For example, synchronizing an Ethereum full node with an archive requires approximately 21,358.246 GB as of March 7, 2025\footnote{https://etherscan.io/chartsync/chainarchive}. Similarly, by early 2024, Solana's blockchain ledger had reportedly exceeded 150 TB\footnote{https://stakin.com/blog/solana-a-2024-ecosystem-overview}.

Effectively processing and analyzing the vast and complex data generated by blockchain networks necessitates advanced analytical techniques. Traditional data analytics, which has been widely employed for decision-making, business intelligence, and predictive modeling (e.g., \cite{moreira2018general, Mafrur2023, sharaf2023efficient}), was originally designed for structured data in relational databases. However, these conventional methods are not well-suited to accommodate the unique characteristics of blockchain data. Unlike structured databases, blockchain data is not only massive but also highly heterogeneous, encompassing diverse transaction types across various blockchain networks~\cite{heo2024blockchain}. 

These complexities necessitate specialized approaches, leading to the development of analytics tools and platforms specifically designed to efficiently manage, process, and interpret blockchain data. However, while these tools have seen significant advancements in industry applications, academic research in this area remains relatively underdeveloped. Based on our research and literature review, we categorize blockchain data analytics tools into several key groups, each reflecting distinct functionalities and applications.

\noindent \textbf{Block Explorers}: These platforms provide a detailed view of blockchain transactions, blocks, and wallet addresses, allowing users to track asset movements and verify transaction histories. They are essential for transparency within blockchain networks.  \textit{Academic Works}: Research contributions include MiningVis~\cite{tovanich2021miningvis}, BitAnalysis~\cite{sun2022bitanalysis}, SilkViser~\cite{zhong2020silkviser}, BIVA~\cite{oggier2018biva}, XBlock~\cite{zheng2020xblock}, and DataEther~\cite{chen2019dataether}, which provide visualization and analytical frameworks for blockchain transactions and mining activities which focus on Bitcoin and Ethereum.  
\textit{Industry Platforms}: Prominent industry platforms include Etherscan~\cite{ref_etherscan}, Blockchain.com~\cite{ref_blockchain}, Blockchair~\cite{ref_blockchair}, BscScan~\cite{ref_bscscan}, and Solscan~\cite{ref_solscan}, offering real-time blockchain transaction tracking.

\noindent \textbf{On-chain Data Providers}: These platforms provide structured access to blockchain data, enabling users to query with/without API and visualize blockchain data.
\textit{Academic Works}: Studies such as Ethanos~\cite{kim2021ethanos}, ChainSync~\cite{zhang2023chainsync}, and EtherNet~\cite{hou2022automating} focus on methodologies for indexing and querying blockchain datasets.
\textit{Industry Platforms}: This category is divided into four types, \textit{Indexers}: The Graph~\cite{ref_thegraph}, Bitquery~\cite{ref_bitquery}, and Covalent~\cite{ref_covalent} enable efficient blockchain data indexing and querying.  \textit{SQL Editors and Visualization Tools}: Dune Analytics~\cite{ref_dune} and Flipside Crypto~\cite{ref_flipside} provide SQL-based querying and visualization capabilities. \textit{No-code Visualization Tools}: Arkham Intelligence~\cite{ref_arkham}, Footprint Analytics~\cite{ref_footprint}, and Dapplooker~\cite{ref_dapplooker} enable non-programmers to visualize blockchain without writing any code or SQL.
        
\noindent \textbf{Research Platforms}: These platforms provide in-depth analytics and insights into blockchain networks, token economies, and market trends, widely used by investors, analysts, and researchers. \textit{Academic Works}: NFTDisk~\cite{wen2023nftdisk}, DenseFlow~\cite{lin2024denseflow}, NFTTracer~\cite{cao2024nftracer}, MindTheDapp~\cite{ibba2024mindthedapp}, NFTTeller~\cite{cao2023nfteller}, and Graphsense~\cite{haslhofer2021graphsense} focus on NFT analytics, transaction tracking, and decentralized application research. \textit{Industry Platforms}: Leading research platforms include Messari~\cite{ref_messari}, Token Terminal~\cite{ref_tokenterminal}, Nansen~\cite{ref_nansen}, Chainalysis~\cite{ref_Chainalysis}, IntoTheBlock~\cite{ref_intotheblock}, Glassnode~\cite{ref_glassnode}, and Bubblemaps~\cite{ref_bubblemap} offering comprehensive market analytics and insights for both individual and enterprise clients.

\noindent \textbf{Market Data Providers}: These platforms aggregate and analyze cryptocurrency market trends, token prices, and trading volumes. \textit{Academic Works}: Research such as Pele et al.~\cite{pele2020statistical} and Wu et al.~\cite{wu2018classification} provides foundational studies on market data aggregation and study comparison among crypto token based on statistical analysis. \textit{Industry Platforms}: Leading market data providers include CoinGecko~\cite{ref_coingecko}, CoinMarketCap~\cite{ref_coinmarketcap}, CryptoCompare~\cite{ref_cryptocompare}, TradingView~\cite{ref_tradingview}, DefiLlama~\cite{ref_defiliama}, DEX Screener~\cite{ref_dexscanner}, and BirdEye~\cite{ref_birdeye}.

Furthermore, this study examines the key challenges in blockchain data analytics, including: (1) data accessibility, (2) scalability, (3) accuracy and (4) interoperability. Our findings highlight the critical need for enhanced collaboration between industry and academia to address these challenges and foster innovation in blockchain analytics. By integrating the expertise and resources of both sectors, the development of more robust tools and methodologies can be accelerated, ultimately unlocking the full potential of blockchain data analytics.

\section{Methodology}

To explore blockchain analytics from both academic papers and practical perspectives. Firstly, we systematically collect academic papers and research articles by searching for relevant keywords, including \textit{blockchain data analytics, crypto analytics, blockchain visualization, bitcoin analytics}, and \textit{blockchain data warehouse}. In addition, for each paper that contained our keywords, we also examined all papers that cited it to ensure we included all relevant works. This ensures comprehensive coverage of existing studies and methodologies in the field. 
Secondly, in addition to reviewing literature, we identify and analyze the existing blockchain analytic tools from the industry. These tools are tested based on their functionalities and practical implementations.
Lastly, we conduct a review and categorize both the collected literature and identified tools based on their functionalities, use cases, and contributions to the field. This categorization aims to provide a structured and clearer understanding of the current landscape and emerging trends in blockchain analytics.

\section{Blockchain Data Analytics}
Blockchain data analytics plays a vital role, particularly in the financial sector (i.e., cryptocurrency space), by enhancing transparency and efficiency within decentralized finance (DeFi) \cite{jiang2023decentralized} and non-fungible tokens (NFTs) \cite{wang2021non}. Through real-time monitoring, risk assessment, and strategic decision-making, blockchain analytics benefits both end users and DeFi protocols, contributing to the development of robust financial ecosystems. 

In traditional database systems, data processing and analysis typically involve constructing a data warehouse using \textit{Online Analytical Processing (OLAP)}. The process begins with extracting, cleaning, transforming, and loading data into the data warehouse, followed by building multidimensional data models with OLAP. OLAP is designed for large-scale data aggregation and complex analytical queries. OLAP supports various multidimentional data operations, including roll-up, drill-down, slice-and-dice, and pivoting (e.g., \cite{chaudhuri1997overview, bohm2016operational}).

On the other hand, \textit{Online Transaction Processing (OLTP)} systems primarily automate clerical data processing tasks, such as e-commerce and banking transactions, which form the backbone of daily transactions. While OLTP prioritizes speed and reliability to ensure transactional integrity, OLAP facilitates deep exploration of historical data, uncovering trends that drive strategic insights.

Traditionally, OLAP and OLTP have been implemented using centralized database systems, where transactional databases serve as the source for analytical data warehouses. However, the emergence of decentralized technologies, particularly blockchain, has introduced new paradigms for large-scale data analytics. Unlike conventional databases, blockchain operates as a distributed ledger, presenting unique challenges and opportunities for data processing and analytics~\cite{vo2018research}.

While blockchain fundamentally differs from traditional database systems, it still involves the storage, retrieval, and processing of transactional data. Blockchain data is stored in cryptographically linked structures, such as Merkle trees~\cite{nakamoto2008bitcoin} (in Bitcoin) or Merkle Patricia Trie~\cite{wood2014ethereum} (in Ethereum), ensuring data integrity and efficient verification. Moreover, blockchain data is typically encoded and requires decoding for meaningful analysis. 
In blockchain analytics, a data warehouse is constructed by retrieving blockchain data (e.g., Ethereum) from nodes such as QuickNode~\cite{quicknode}, Alchemy~\cite{alchemy}, Chainstack~\cite{chainstack}, and Infura~\cite{infura}. The retrieved data, which is initially encoded, undergoes a decoding process before being processed through an Extract, Transform, and Load (ETL) pipeline. This pipeline structures blockchain data into a structured database, which is then stored in a centralized database system for analytics~\cite{zheng2020xblock}.
In the next section, we review existing blockchain data analytics tools from both academic research and industries.

\section{Review of Blockchain Data Analytics Tools}
In this paper, we summarize our review of both academic research and existing industry applications of blockchain data analytics tools, as presented in Table~\ref{table1}. These tools are categorized into four main types: block explorers, on-chain data providers, research platforms, and market data providers, which are listed horizontally on the left side of the table.

To enhance clarity, we introduce additional categorizations through columns. The first categorization is based on the type of work, distinguishing between academic research and industry applications. The second categorization considers supported blockchains, which include Bitcoin, Ethereum Virtual Machine (EVM)-based blockchains (e.g., Ethereum and its derivatives), non-EVM blockchains (e.g., Solana, Sui, and others), and legacy-chain platforms that support a diverse range of blockchains beyond Bitcoin, EVM, and non-EVM ecosystems (e.g., Dogecoin, Ripple, Zcash, Litecoin, other coins in Bitcoin era).

Another key metric we incorporate is the target user, classified into novice, intermediate, and advanced levels. Novice users can easily interpret insights directly from the visualized data presented by the tool without requiring technical expertise. Intermediate users typically need to write SQL queries to retrieve specific blockchain data and generate insights. Advanced users, such as those using (e.g., The Graph~\cite{ref_thegraph}, Bitquery~\cite{ref_bitquery}), must interact with an API to fetch blockchain data, which they then process and visualize using external tools, such as Python visualization libraries, to extract meaningful insights.

Finally, we analyze the types of visualizations provided by these tools, as visualization plays a critical role in blockchain data analytics. The visualization types are further categorized into subtypes: bar charts and time series charts, which are the most commonly used; graphs and Sankey diagrams, which are particularly useful for tracking money flows; and mixed visualizations, where tools integrate multiple visualization types.

{\scriptsize
\begin{longtable}{|c|l|c|c|c|c|c|c|c|c|c|c|c|c|c|c|}
    \caption{Classification Table of Blockchain Data Analytics Tools} \\
    \hline
    \multirow{2}{*}{\rotatebox{90}{\textbf{\centering Category}}} & \multirow{2}{*}{\textbf{Tool Name}} & \multicolumn{2}{c|}{\textbf{Type}} & \multicolumn{4}{c|}{\textbf{Blockchain}} & \multicolumn{3}{c|}{\textbf{Target User}} & \multicolumn{5}{c|}{\textbf{Visualization Type}} \\
    \cline{3-16}
    & & \rotatebox{90}{Industry} & \rotatebox{90}{Academic paper} & \rotatebox{90}{Bitcoin} & \rotatebox{90}{Ethereum/EVM} & \rotatebox{90}{Non-EVM} & \rotatebox{90}{Legacy-Chain} & \rotatebox{90}{Novice} & \rotatebox{90}{Intermediate} & \rotatebox{90}{Advanced} & \rotatebox{90}{Bar Charts} & \rotatebox{90}{Time Series} & \rotatebox{90}{Graph} & \rotatebox{90}{Sankey diag.} & \rotatebox{90}{Mixed} \\
    \hline
    
    \endfirsthead
    
    \multicolumn{16}{c}{{Continued from previous page}} \\
    \hline
    \multirow{2}{*}{\rotatebox{90}{\textbf{Category}}} & \multirow{2}{*}{\textbf{Tool Name}} & \multicolumn{2}{c|}{\textbf{Type}} & \multicolumn{4}{c|}{\textbf{Blockchain}} & \multicolumn{3}{c|}{\textbf{Target User}} & \multicolumn{5}{c|}{\textbf{Visualization Type}} \\
    \cline{3-16}
    & & \rotatebox{90}{Industry} & \rotatebox{90}{Academic paper} & \rotatebox{90}{Bitcoin} & \rotatebox{90}{Ethereum/EVM} & \rotatebox{90}{Non-EVM} & \rotatebox{90}{Legacy-Chains} & \rotatebox{90}{Novice} & \rotatebox{90}{Intermediate} & \rotatebox{90}{Advanced} & \rotatebox{90}{Bar Charts} & \rotatebox{90}{Time Series} & \rotatebox{90}{Graph} & \rotatebox{90}{Sankey diag.} & \rotatebox{90}{Mixed} \\
    \hline
    \endhead
    
    \hline \multicolumn{16}{|r|}{{Continued on next page}} \\ \hline
    \endfoot
    
    \hline
    \endlastfoot
    
    \multirow{5}{*}{\rotatebox{90}{\parbox{2cm}{\centering Block \\ Explorer}}}

    & Blockchain.com~\cite{ref_blockchain}    & x  &    & x  & x  &    & x  & x  &    &    & x  & x  &    &    & x  \\
    & Blockchair~\cite{ref_blockchair}        & x  &    & x  & x  & x  & x  & x  &    &    & x  & x  &    &    & x  \\
    & Etherscan~\cite{ref_etherscan}          & x  &    &    & x  &    &    & x  &    &    &    & x  &    &    & x  \\
    & BscScan~\cite{ref_bscscan}              & x  &    &    & x  &    &    & x  &    &    &    & x  &    &    & x  \\
    & Solscan~\cite{ref_solscan}              & x  &    &    &    & x  &    & x  &    &    & x  & x  &    &    & x  \\
    & MiningVis~\cite{tovanich2021miningvis}  &    & x  & x  &    &    &    & x  &    &    & x  & x  &    &    & x  \\ 
    & BitAnalysis~\cite{sun2022bitanalysis}   &    & x  & x  &    &    &    & x  &    &    & x  &    & x  & x  & x  \\
    & BIVA~\cite{oggier2018biva}              &    & x  & x  &    &    &    & x  &    &    &    &    & x  &    &    \\
    & SilkViser~\cite{zhong2020silkviser}     &    & x  & x  & x  &    &    & x  & x  & x  & x  &    &    & x  & x  \\
    & DataEther~\cite{chen2019dataether}      &    & x  &    & x  &    &    & x  &    &    &    & x  &    &    &    \\
    & XBlock~\cite{zheng2020xblock}           &    & x  &    & x  &    &    & x  &    &    & x  & x  &    &    & x  \\

    \hline
  \multirow{5}{*}{\raisebox{-.5\height}{\rotatebox{90}{\parbox{2cm}{\centering On-Chain \\ Data Provider}}}}

    & The Graph~\cite{ref_thegraph}           & x  &    &    & x  &    &    &    &    & x  &    &    &    &    &    \\
    & Bitquery~\cite{ref_bitquery}            & x  &    & x  & x  & x  & x  &    &    & x  &    &    &    &    &    \\
    & Covalent~\cite{ref_covalent}            & x  &    & x  & x  & x  & x  &    &    & x  &    &    &    &    &    \\
    & Dune Analytics~\cite{ref_dune}          & x  &    & x  & x  & x  & x  &    & x  & x  & x  & x  &    &    & x  \\
    & Flipside Crypto~\cite{ref_flipside}     & x  &    & x  & x  & x  &    &    & x  & x  & x  & x  &    &    & x  \\
    & Arkham Intelligence~\cite{ref_arkham}   & x  &    & x  & x  & x  & x  &    &    & x  & x  & x  &    &    & x  \\
    & Footprint Analytics~\cite{ref_footprint}& x  &    &    & x  & x  &    & x  &    &    & x  & x  &    &    & x  \\
    & DappLooker~\cite{ref_dapplooker}        & x  &    &    & x  &    &    & x  &    &    & x  & x  &    &    & x  \\
    & Ethanos~\cite{kim2021ethanos}           &    & x  &    & x  &    &    &    &    & x  &    &    &    &    &    \\
    & ChainSync~\cite{zhang2023chainsync}     &    & x  &    & x  &    &    &    &    & x  &    &    &    &    &    \\
    & EtherNet~\cite{hou2022automating}       &    & x  &    & x  &    &    &    &    & x  &    &    & x  &    &    \\

    \hline
   \multirow{5}{*}{\rotatebox{90}{\parbox{2cm}{\centering Research \\ Platform}}}

    & Messari~\cite{ref_messari}                    & x  &    & x  & x  & x  & x  & x  &    &    & x  & x  &    &    & x  \\
    & Nansen~\cite{ref_nansen}                      & x  &    & x  & x  & x  & x  & x  &    &    & x  & x  &    & x  & x  \\
    & Token Terminal~\cite{ref_tokenterminal}       & x  &    & x  & x  & x  & x  & x  &    &    & x  & x  & x  &    & x  \\
    & Chainalysis~\cite{ref_Chainalysis}            & x  &    & x  & x  & x  & x  & x  & x  &    & x  & x  & x  & x  & x  \\
    & IntoTheBlock~\cite{ref_intotheblock}          & x  &    & x  & x  & x  & x  & x  &    &    & x  & x  &    &    & x  \\
    & Glassnode~\cite{ref_glassnode}                & x  &    & x  & x  & x  & x  & x  &    &    & x  & x  &    &    & x  \\
    & Bubblemaps~\cite{ref_bubblemap}               & x  &    &    & x  &    &    & x  &    &    &    &    & x  &    &    \\
    & Graphsense~\cite{haslhofer2021graphsense}     &    & x  & x  &    &    & x  & x  &    &    &    &    & x  & x  &    \\
    & BlockSci~\cite{kalodner2020blocksci}          &    & x  & x  &    &    & x  &    &    & x  & x  & x  & x  &    & x  \\
    & DenseFlow~\cite{lin2024denseflow}             &    & x  &    & x  &    &    &    &    & x  &    &    & x  &    &    \\
    & MindTheDApp~\cite{ibba2024mindthedapp}        &    & x  &    & x  &    &    &    &    & x  &    &    & x  &    & x  \\
    & NFTTeller~\cite{cao2023nfteller}              &    & x  &    & x  &    &    & x  &    &    & x  & x  &    &    & x  \\
    & NFTTracer~\cite{cao2024nftracer}              &    & x  &    & x  &    &    & x  &    &    & x  & x  & x  &    & x  \\
    & NFTDisk~\cite{wen2023nftdisk}                 &    & x  &    & x  &    &    & x  &    &    &    &    & x  & x  &    \\

    \hline
   \multirow{5}{*}{\rotatebox{90}{\parbox{3cm}{\centering Market \\ Data Provider}}}

    & CoinGecko~\cite{ref_coingecko}            & x  &    & x  & x  & x  & x  & x   &    &    & x  & x  &    &    & x  \\
    & CoinMarketCap~\cite{ref_coinmarketcap}    & x  &    & x  & x  & x  & x  & x   &    &    & x  & x  &    &    & x  \\
    & CryptoCompare~\cite{ref_cryptocompare}    & x  &    & x  & x  & x  & x  & x   &    &    & x  & x  &    &    & x  \\
    & TradingView~\cite{ref_tradingview}        & x  &    & x  & x  & x  & x  & x   &    &    & x  & x  &    &    & x  \\
    & DefIllama~\cite{ref_defiliama}            & x  &    & x  & x  & x  &    & x   &    &    & x  & x  &    &    & x  \\
    & DEX Screener~\cite{ref_dexscanner}        & x  &    &    & x  & x  &    & x   &    &    & x  & x  &    &    & x  \\
    & BirdEye~\cite{ref_birdeye}                & x  &    &    & x  & x  &    & x   &    &    & x  & x  &    &    & x  \\
    & Pele et al.~\cite{pele2020statistical}    &    & x  & x  & x  & x  & x  & x   &    &    & x  & x  & x  & x  & x  \\
    & Wu et al.~\cite{wu2018classification}     &    & x  & x  & x  & x  & x  & x   &    &    & x  & x  & x  & x  & x  \\
    
    \hline
\end{longtable}
\label{table1}
}

\subsection{Block Explorers}
Blockchain explorers serve as essential tools for analyzing on-chain activities, providing real-time insights into transactions, blocks, addresses, and network behavior. These tools act as an interface between users and blockchain ledgers, enabling transparent access to blockchain data. In academic research, several blockchain exploration frameworks have been proposed to enhance data visualization, transaction analysis, and network clustering.

Several studies have contributed to this domain. BitAnalysis~\cite{sun2022bitanalysis} equips law enforcement and regulatory agencies with advanced tools for visualizing and analyzing Bitcoin transactions. By identifying wallet clusters and tracking fund movements, its clustering algorithms reveal wallet correlations, while visualization techniques, such as the connection diagram and Bitcoin flow map, enhance transaction monitoring and forensic analysis.

BIVA~\cite{oggier2018biva} enables data exploration and subgraph visualization around key nodes. It incorporates a general flow-based clustering algorithm for directed graphs alongside specialized Bitcoin network techniques for wallet address aggregation. These capabilities provide deeper insights into blockchain network structures and transaction patterns.

SilkViser~\cite{zhong2020silkviser} introduces a user-centric visualization approach designed to simplify the understanding of cryptocurrency transaction systems. By leveraging intuitive visual representations, it facilitates novice users in grasping fundamental concepts while empowering experienced users with efficient transaction data analysis. This approach enhances accessibility and paves the way for further advancements in cryptocurrency visualization tools.

MiningVis~\cite{tovanich2021miningvis} offers a structured approach for analyzing the Bitcoin mining ecosystem. It provides insights into market trends, mining pool rankings, and behavioral patterns by integrating external data sources, such as mining pool attributes and Bitcoin-related news. Through this contextual analysis, MiningVis contributes to a more comprehensive understanding of the mining landscape.

DataEther~\cite{chen2019dataether} focuses on various applications, such as profiling account balances, analyzing reentrancy attacks, and identifying non-deployable contracts, offering new insights into Ethereum’s ecosystem and security risks.  
XBlock-ETH~\cite{zheng2020xblock} is a data processing framework designed to extract, transform, and analyze Ethereum blockchain data. It processes raw blockchain data into structured statistical datasets, enabling comprehensive exploration and analysis. This approach streamlines Ethereum data processing, making it more accessible for research and analytics.

Beyond academic research, several widely used industry blockchain explorers provide access to blockchain data. Blockchair~\cite{ref_blockchair} and Blockchain.com~\cite{ref_blockchain} support multiple blockchain networks, including Bitcoin, Ethereum, and various EVM and non-EVM chains. Etherscan~\cite{ref_etherscan} specializes in Ethereum and its dapps, while BscScan~\cite{ref_bscscan} is Binance Smart Chain explorer. Similarly, Solscan~\cite{ref_solscan} serves as the primary blockchain explorer for Solana.

\subsection{On-Chain Data Providers}

On-chain data providers are essential for structuring and indexing blockchain data, transforming raw on-chain information into accessible, queryable formats for analysis. Several studies have advanced this field by developing frameworks for blockchain data extraction, transformation, and storage. 

ETHANOS~\cite{kim2021ethanos} introduces a lightweight bootstrapping mechanism for Ethereum, significantly reducing synchronization costs by periodically sweeping inactive accounts and omitting the download of outdated transactions. ChainSync~\cite{zhang2023chainsync} presents a multi-chain ETL (Extract, Transform, Load) system designed to streamline multiple blockchains data processing. EtherNet~\cite{hou2022automating} automates blockchain ETL processes on Google BigQuery. It generates graph representations of Ethereum's transaction network, enabling efficient visualization and mining of blockchain interactions at scale. This framework enhances blockchain analytics by providing a structured method for exploring Ethereum transaction history.

Beyond academic research, industry platforms provide solutions for indexing, querying, and visualizing blockchain data. Some platforms specialize in blockchain data indexing and querying, such as The Graph~\cite{ref_thegraph}, Bitquery~\cite{ref_bitquery}, and Covalent~\cite{ref_covalent}, which support API use cases for dApp development. Others focus on SQL-based querying and data visualization, with platforms like Dune Analytics~\cite{ref_dune} and Flipside Crypto~\cite{ref_flipside} enabling users to execute SQL query on the blockchain data and generate visualizations. Additionally, no-code visualization tools such as Arkham Intelligence~\cite{ref_arkham}, Footprint Analytics~\cite{ref_footprint}, and Dapplooker~\cite{ref_dapplooker} provide intuitive interfaces that allow users to explore and analyze blockchain data without requiring programming or SQL expertise. These industry solutions enhance accessibility to blockchain data, catering to a diverse range of users from developers and analysts to regulators and non-technical users.

\subsection{Blockchain Research Platforms}
Several studies have contributed to this field by developing innovative frameworks for blockchain research tools. GraphSense~\cite{haslhofer2021graphsense} employs graph algorithms to facilitate interactive investigations of monetary flows, enhancing transparency in blockchain transactions. BlockSci~\cite{kalodner2020blocksci}, an open-source blockchain research platform, supports multiple blockchain networks and leverages an in-memory analytical database, significantly improving query efficiency over traditional graph databases. DenseFlow~\cite{lin2024denseflow} introduces a framework for detecting and tracing money laundering activities by identifying dense subgraphs, thereby strengthening illicit activity detection mechanisms. MindTheDApp~\cite{ibba2024mindthedapp} focuses on the structural analysis of Ethereum-based decentralized applications (DApps), utilizing a complex network-driven approach to offer deeper insights into DApp interactions. 

NFTeller~\cite{cao2023nfteller}, NFTracer~\cite{cao2024nftracer}, and NFTDisk~\cite{wen2023nftdisk} are visual analytics systems designed to enhance the understanding of NFT market dynamics. NFTeller employs a dual-centric workflow that incorporates static and dynamic impact attributes to assess NFT market performance. NFTracer focuses on analyzing transaction-flow substitutive systems, revealing how NFT projects with high similarity tend to replace one another and how factors such as stakeholder influx and project freshness influence market positioning. NFTDisk, on the other hand, provides a visualization system for detecting wash trading in NFT markets by combining a radial overview of transactions with a flow-based module for detailed analysis. 

\begin{figure}[htbp]
    \centering
    \includegraphics[width=0.8\textwidth]{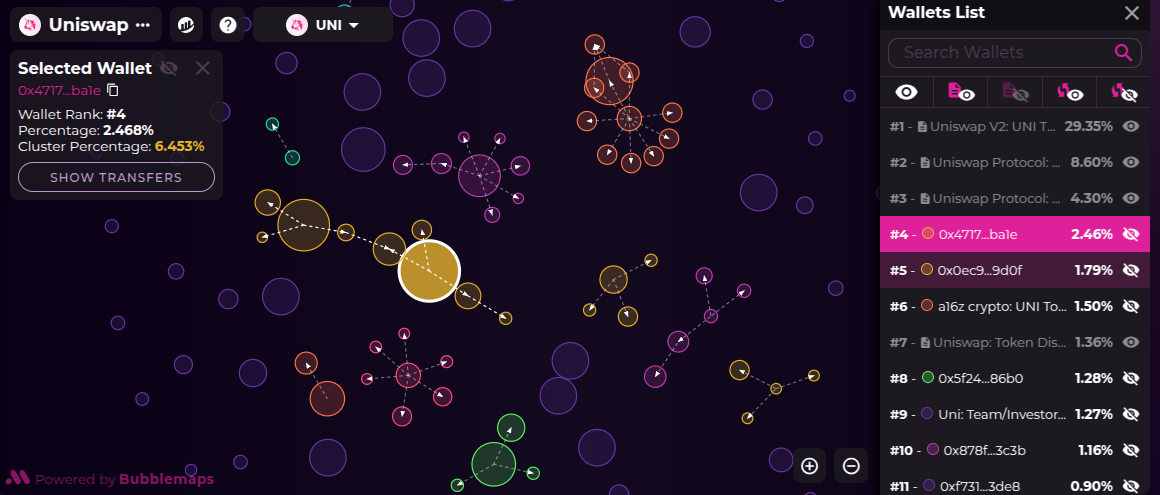}
    \caption{Bubblemap visualization}
    \label{fig:bubblemap}
\end{figure}

In addition to academic research, industry-driven blockchain intelligence platforms provide essential analytics for a broad range of users, including individual investors, institutional clients, regulatory bodies, and financial analysts. Messari~\cite{ref_messari} offers proprietary analytics dashboards that track blockchain projects, DeFi protocols, and token performance, supplemented by AI-powered sentiment analysis for assessing market trends and investor sentiment. Additionally, it provides institutional-grade research reports on emerging blockchain technologies, governance models, and regulatory developments. Nansen~\cite{ref_nansen} specializes in tracking smart money flows, offering insights into the movements of institutional and high-net-worth investors, as well as on-chain behavior patterns of individual traders. Chainalysis~\cite{ref_Chainalysis} is a leading platform for blockchain forensic analysis, focusing on fraud detection, anti-money laundering (AML) compliance, and tracking illicit financial activities across blockchain networks. Token Terminal~\cite{ref_tokenterminal} provides fundamental financial data on crypto projects, analyzing revenue generation, protocol fees, and user adoption metrics. IntoTheBlock and Glassnode offer blockchain-based trading analytics, equipping investors with on-chain indicators, trading signals, and predictive models to optimize investment strategies. Bubblemaps~\cite{ref_bubblemap} utilizes a graph-based visualization approach to represent token holder distributions, where wallets are displayed as interconnected bubbles, allowing users to track ownership concentration and potential market manipulation as shown in Figure~\ref{fig:bubblemap}. These platforms collectively enhance blockchain intelligence by providing valuable tools for market research, financial analysis, and regulatory compliance.

\subsection{Market Data Providers}

Market data providers aggregate real-time and historical price data, trading volumes, and liquidity metrics from both centralized exchanges (CEXs) and decentralized exchanges (DEXs). These platforms serve retail and institutional traders, portfolio managers, and financial analysts by enabling data-driven investment decisions. Notable providers such as CoinGecko~\cite{ref_coingecko} and CoinMarketCap~\cite{ref_coinmarketcap}, CryptoCompare~\cite{ref_cryptocompare}, DeFiLlama~\cite{ref_defiliama}. In addition, DEX-specific aggregators like DEX Screener~\cite{ref_dexscanner} and BirdEye~\cite{ref_birdeye} concentrate solely on decentralized exchange markets, offering real-time tracking of liquidity pools, trading volumes, and token swaps.

Academic research has also contributed to understanding market dynamics and asset classification in the crypto space. Pele et al.~\cite{pele2020statistical} analyze key statistical factors that differentiate cryptocurrencies from traditional financial assets. By applying classification techniques to daily log-return time series, the study characterizes assets through multidimensional statistical vectors incorporating variance, skewness, kurtosis, tail probability, etc. Wu et al.~\cite{wu2018classification} examine the market capitalization distributions of cryptocurrencies, demonstrating that coins and tokens follow power-law distributions with distinct tail exponent values. Their findings suggest that cryptocurrencies exhibit proportional growth patterns similar to firms, cities, and web pages, with tokens representing a highly dynamic and rapidly evolving ecosystem, largely influenced by Initial Coin Offerings (ICOs). The study predicts that, over time, token markets will stabilize as technological innovations drive sustainable adoption.

\section{Challenges in Blockchain Data Analytics}

Blockchain data analytics faces several challenges that hinder its effectiveness and efficiency. These challenges include accessibility, scalability, interoperability, and accuracy. In this section, we provide an in-depth discussion of each challenge.

\subsection{Accessibility}

Blockchain operates as a distributed ledger, where network participants (i.e., miners or validators) validate transactions and maintain data integrity. For example, Bitcoin relies on a decentralized network of miners who validate transactions through Proof-of-Work (PoW), while Ethereum initially used PoW but has since transitioned to a Proof-of-Stake (PoS) mechanism. In Ethereum's PoS model, participants who stake a minimum of 32 ETH~\cite{ref_eth_stake} can set up their own validator nodes to be a validator. To access blockchain data, users must either run a full node and synchronize historical data using such as the Geth client application or rely on third-party node service providers such as QuickNode~\cite{quicknode}, Alchemy~\cite{alchemy}, Infura~\cite{infura}, and Chainstack~\cite{chainstack}, which offer scalable and reliable blockchain infrastructure for data access and querying.

A significant accessibility challenge lies in the raw and unstructured nature of blockchain data. Public blockchain networks such as Bitcoin and Ethereum store transactional data in low-level formats, requiring additional processing before meaningful analysis can be conducted. For instance, smart contract interactions on Ethereum must be decoded using the Contract Application Binary Interface (ABI) to retrieve structured data. To simplify this process, indexing services such as The Graph~\cite{ref_thegraph} provide structured APIs that enable efficient data retrieval, eliminating the need for direct decoding. However, these platforms primarily focus on specific ecosystems, such as EVM-compatible chains, which limits their applicability to non-EVM blockchains. Other platforms, such as Dune Analytics~\cite{ref_dune}, offer broader multi-chain support and allow users to query blockchain data using SQL. However, executing complex SQL queries can be time-consuming, and in some cases, users may need to upgrade to premium services to improve query performance. An alternative approach is to access blockchain datasets available through Google BigQuery\footnote{https://cloud.google.com/blockchain-analytics/docs/supported-datasets}, which provides structured blockchain data for analysis using SQL-based queries. For users who prefer self-hosted solutions, blockchain data can be processed using open-source ETL (Extract, Transform, Load) frameworks such as the Blockchain ETL project\footnote{https://github.com/blockchain-etl}, which facilitates structured extraction from raw blockchain data. 

Overall, blockchain data accessibility has improved significantly, with services ranging from raw data providers (e.g., QuickNode~\cite{quicknode}, Alchemy~\cite{alchemy}, Infura~\cite{infura}, and Chainstack~\cite{chainstack}) to advanced analytics platforms such as Messari~\cite{ref_messari} and Nansen~\cite{ref_nansen}, which offer ready-to-use insights. However, the most reliable and high-performance data services often require a subscription, limiting access for researchers with constrained resources. Another key challenge is enhancing the interpretability of blockchain data, particularly for beginners. Advanced data visualization techniques, such as those proposed by SilkViser~\cite{zhong2020silkviser}, can improve the presentation of blockchain data, making it more intuitive for users with varying levels of expertise. Additionally, automated visualization recommendation systems, such as those proposed by Mafrur et al.~\cite{Mafrur2023, sharaf2023efficient, mafrur2018dive}, can automatically extract and recommend meaningful insights from blockchain data. Future research should focus on developing more effective visualization methods to enhance blockchain data accessibility for beginners, intermediate users, and advanced analysts alike.

\subsection{Scalability}

Scalability remains a major challenge in blockchain data analytics due to the immense volume of blockchain records. As of March 7, 2025, synchronizing an Ethereum full node with an archive requires approximately 21,358.246 GB, while by early 2024, Solana’s blockchain ledger had reportedly exceeded 150 TB. Extracting, transforming, and loading (ETL) blockchain data into structured formats is essential for efficient querying, typically using tools such as Data Build Tool (DBT)~\cite{ref_dbt} for data pipelining. The processed data is then stored in Online Analytical Processing (OLAP) systems, which are optimized for analytical workloads by using column-based storage. OLAP databases are particularly effective for executing complex aggregation queries, such as calculating active addresses, measuring user retention, or determining protocol revenue. However, despite the advantages of OLAP, scalability remains a significant issue as blockchain datasets continue to expand, requiring more efficient query execution strategies.

To address scalability challenges, several approaches have been developed, including parallel query execution techniques, as demonstrated by Kılıç et al.~\cite{kilicc2022parallel}. Open-source distributed SQL query engines such as ClickHouse~\cite{schulze2024clickhouse} and Trino~\cite{fuller2022trino} are widely used to enhance scalability in OLAP-based blockchain analytics. ClickHouse is an open-source OLAP database optimized for high-performance analytics over large-scale datasets with high ingestion rates. It employs a log-structured merge (LSM) tree-based storage format combined with background data transformations, such as aggregation and archiving, to improve query efficiency. Its vectorized execution engine and advanced pruning techniques further contribute to its exceptional performance, making it one of the fastest analytical databases available. Similarly, Trino~\cite{fuller2022trino} is a distributed SQL query engine designed for querying large datasets across multiple heterogeneous data sources. Originally developed as a fork of Presto~\cite{sethi2019presto}, Trino has evolved independently, with both systems continuing to develop distinct features and optimizations.

Several blockchain analytics platforms have integrated these scalable query engines to handle large blockchain datasets effectively. Dune Analytics~\cite{ref_dune} utilizes Trino as its query engine, structuring blockchain data into separate schemas for different blockchain networks, even within the same blockchain family, such as Ethereum and Binance Smart Chain (BSC). In contrast, Covalent~\cite{ref_covalent} employs ClickHouse to maintain a unified blockchain schema, consolidating data from multiple networks into a single database for more streamlined querying. Other platforms leverage enterprise cloud solutions to address scalability concerns. Nansen~\cite{ref_nansen}, for example, utilizes Google BigQuery, while Flipside Crypto~\cite{ref_flipside} relies on Snowflake, both of which are well-known for their high-performance query capabilities. 

Academic research, such as Ethanos~\cite{kim2021ethanos}, explores bootstrapping methods to improve Ethereum node synchronization. Currently, Ethereum full nodes, including archive nodes, require terabytes of storage, making it impractical for ordinary clients using PCs to sync. Ethanos addresses this issue by optimizing Ethereum’s state trie, periodically removing inactive accounts and retaining only active ones from recent transactions. This method allows nodes to bootstrap with a significantly smaller state trie while still verifying all historical transactions, reducing storage overhead. Evaluations on real Ethereum data show that Ethanos maintains the state trie size within a few hundred MB, preventing unbounded growth while requiring minimal restore transactions. 

Overall, both academia and industry have made significant efforts to improve the scalability of blockchain data processing. However, challenges remain, particularly when executing complex queries on large datasets. For instance, even though Dune Analytics utilizes the Trino query engine, query execution times can still be an issue when joining multiple schemas. An alternative academic approach involves using sampling or bootstrapping techniques instead of processing the entire dataset, allowing for the use of margin of error estimations in query results~\cite{hellerstein1997online}. This method can be useful when absolute precision is not required, providing a scalable solution for gaining broader insights into blockchain data.

\subsection{Accuracy}

As shown in Table~\ref{table1}, both academic research and industry blockchain analytics tools offer a wide range of solutions, from processing raw blockchain data to executing SQL queries or simply consuming pre-analyzed insights. However, a fundamental question arises: how can we verify the accuracy of these outputs? Are the insights generated by these platforms reliable, and who is responsible for their validation? Blockchain data is often subject to inconsistencies due to issues such as data duplication, erroneous smart contracts, and incomplete transaction records. Additionally, user-generated queries on open analytics platforms may introduce biases or inaccuracies, impacting the reliability of insights. As highlighted by ~\cite{mafrur2020quality, mafrur2023vizput}, data quality plays a crucial role in visual analytics, directly affecting the accuracy and interpretability of analytical results. Ensuring the integrity of blockchain data analytics is essential for deriving meaningful insights and supporting informed decision-making.

Another challenge stems from the anonymity of blockchain wallets. In public blockchains such as Bitcoin and Ethereum, all transactions are transparent and publicly recorded; however, unless wallet owners voluntarily disclose their identities, it remains impossible to determine the true ownership of a wallet. Some platforms, such as Arkham Intelligence~\cite{ref_arkham}, attempt to address this issue by offering bounties for wallet labeling. Additionally, Arkham employs an AI-driven system called Ultra, which algorithmically labels wallets based on transaction history and money flow. Other platforms, such as Flipside Crypto, rely on crowdsourced wallet tagging, allowing users to contribute wallet labels collectively. However, these approaches raise concerns about accuracy, as incorrectly labeled wallets can significantly impact analytical insights and lead to misleading conclusions.

When using platforms like Dune Analytics~\cite{ref_dune}, it is advisable to cross-validate results with other sources, such as Token Terminal~\cite{ref_tokenterminal} or DeFiLlama~\cite{ref_defiliama}, to ensure data consistency. Metrics available in one platform should be verified against others to identify potential discrepancies. Additionally, academic research still faces significant challenges in improving blockchain data accuracy, including the development of robust AI algorithms for wallet labeling, optimizing crowdsourcing mechanisms for wallet tagging, and establishing reliable methods for cross-platform data validation. Addressing these challenges will be crucial for enhancing the reliability and trustworthiness of blockchain analytics.

\begin{table}[h!]
\centering
\renewcommand{\arraystretch}{1.2}  
{\scriptsize 
\begin{tabular}{|l|l|l|}  
\hline
\textbf{Blockchain} & \textbf{Type} & \textbf{Description} \\
\hline
Bitcoin~\cite{nakamoto2008bitcoin} & Legacy L1 & First cryptocurrency, UTXO model, PoW. \\
Ethereum~\cite{buterin2013ethereum} & EVM L1 & Smart contracts, DeFi, NFTs, PoW → PoS. \\
BSC~\cite{binance2019bsc} & EVM L1 & Fast, cheap EVM chain by Binance, PoA/DPoS. \\
Avalanche~\cite{rocket2018avalanche} & EVM L1 & High-speed, scalable, PoS, EVM-compatible. \\
Polygon PoS\cite{polygon2023pos} & EVM L1 Sidechain & Fast, low-cost PoS chain, independent security. \\
Solana~\cite{yakovenko2018solana} & Non-EVM L1 & High-speed, Rust contracts, Proof of History. \\
Cardano~\cite{wood2017cardano} & Non-EVM L1 & PoS, UTXO model, Plutus smart contracts. \\
Aptos~\cite{aptos2022aptos} & Non-EVM L1 & High-speed PoS, Move smart contracts. \\
Sui~\cite{sui2022sui} & Non-EVM L1 & Object-based PoS, Move smart contracts. \\
Cosmos~\cite{kwon2019cosmos} & Non-EVM L0 & Interoperability-focused, Tendermint BFT. \\
Polkadot Relay~\cite{wood2017polkadot} & Non-EVM L0 & Secures parachains, WASM runtime. \\
Polkadot Parachains~\cite{wood2017polkadot} & Mixed L1 & Independent chains, some EVM-compatible. \\
Lightning Network~\cite{poelstra2015lightning} & Legacy L2 & Bitcoin L2 for instant low-fee payments. \\
Polygon zkEVM~\cite{polygon2023zkevm} & EVM L2 & Ethereum-compatible zk-Rollup for scaling. \\
Arbitrum~\cite{kalodner2018arbitrum}/Optimism~\cite{optimism2021} & EVM L2 & Off-chain scaling, Rollups. \\
Litecoin~\cite{litecoin2011litecoin} & Legacy L1 & Bitcoin fork, Scrypt PoW, 4× faster blocks. \\
Dogecoin~\cite{dogecoin2013dogecoin} & Legacy L1 & Meme coin, fork of Litecoin, unlimited supply. \\
XRP~\cite{ripple2012ripple} & Legacy L1 & Payment-focused, no mining, fast finality. \\
Zcash~\cite{hopwood2016zcash} & Legacy L1 & Bitcoin fork with zk-SNARK privacy. \\
\hline
\end{tabular}
} 
\caption{Blockchain classification: L0 (Layer 0) provides security and interoperability for connected chains.  
L1 (Layer 1) is a base blockchain with independent security and consensus.  
L2 (Layer 2) is built on top of an L1 to enhance scalability, reducing costs and congestion.  
Non-EVM chains use different execution environments, requiring unique development tools.  
Legacy refers to Bitcoin-era blockchains that lack modern smart contract capabilities but remain significant.}
\label{tab:blockchain_classification}
\end{table}

\subsection{Interoperability}

Interoperability remains one of the biggest challenges in blockchain analytics. The growing number of public blockchain networks introduces fragmentation, as each network operates with its own execution environment and protocol layer. Table~\ref{tab:blockchain_classification} presents a comparison of major existing public blockchains, categorizing them based on their execution environments and architectural layers.

EVM-based blockchains, such as Ethereum~\cite{buterin2013ethereum}, Binance Smart Chain~\cite{binance2019bsc}, and Avalanche~\cite{rocket2018avalanche}, utilize the Ethereum Virtual Machine (EVM) for smart contract execution, ensuring compatibility across the Ethereum ecosystem. In contrast, non-EVM blockchains, including Solana~\cite{yakovenko2018solana}, Polkadot~\cite{wood2017polkadot}, and Cosmos~\cite{kwon2019cosmos}, employ distinct virtual machines and programming languages, prioritizing performance and scalability over EVM compatibility. Consequently, these differences extend to their underlying data structures, making cross-chain data analytics inherently complex.

Public blockchains also differ in their structural hierarchy. Layer-1 (L1) blockchains, such as Bitcoin, Ethereum, and Solana, serve as base networks that independently secure and process transactions using consensus mechanisms like Proof of Work (PoW) or Proof of Stake (PoS). Layer-2 (L2) solutions, including Ethereum’s Optimistic and zk-Rollups, as well as Bitcoin’s Lightning Network, enhance scalability by processing transactions off-chain and periodically settling results back on L1. Unlike L2s, sidechains function as independent L1s that bridge to another blockchain but maintain separate security, highlighting the distinction between interoperability-focused architectures and pure scalability enhancements.

To address interoperability, Layer-0 (L0) solutions like Polkadot~\cite{wood2017polkadot} and Cosmos~\cite{kwon2019cosmos} enable communication between sovereign L1 blockchains. Polkadot’s Relay Chain offers shared security for parachains, facilitating cross-chain messaging via XCMP/XCM protocols. Similarly, Cosmos Hub leverages the Inter-Blockchain Communication (IBC) protocol to connect independent blockchains, though each chain is responsible for its own security. 

Beyond these architectures, legacy blockchain networks continue to play a significant role in digital payments and cryptographic advancements. Litecoin~\cite{litecoin2011litecoin} introduced faster transactions through Scrypt PoW and reduced block times. Dogecoin~\cite{dogecoin2013dogecoin}, a fork of Litecoin, adopted an inflationary model and gained traction as a meme-driven currency. Ripple’s XRP Ledger~\cite{ripple2012ripple} replaced traditional mining with RPCA consensus, optimizing for fast, low-cost transactions. Zcash~\cite{hopwood2016zcash} pioneered privacy-focused blockchain technology using zk-SNARKs, enabling fully shielded transactions.

The fragmentation of blockchain networks, each with unique consensus mechanisms, data structures, and governance models, creates significant interoperability challenges. This issue extends to blockchain analytics, where unified data processing remains difficult due to structural differences between networks. Platforms like Covalent~\cite{ref_covalent} attempt to standardize blockchain data under a single schema for EVM-compatible chains, but non-EVM blockchains like Solana and Cosmos require separate schemas, increasing complexity. As blockchain ecosystems continue to evolve, future networks may introduce even greater architectural diversity, making cross-chain analytics an ongoing challenge, particularly for the blockchain data analytics industry.

\section{Conclusion}

This paper presents literature review of blockchain analytics tools in academic research and evaluates existing blockchain analytics solutions in the industry. Our findings indicate a significant gap between academic advancements and industry developments, which is expected given the rapid evolution of the cryptocurrency space. The blockchain ecosystem has witnessed continuous waves of innovation, including the rise of Initial Coin Offerings (ICOs), Decentralized Finance (DeFi), Non-Fungible Tokens (NFTs), Layer 2 scaling solutions, Real-World Asset (RWA) tokenization, and AI-driven crypto assets. While academic research often lags behind these fast-paced industry trends, we also observe instances where academic contributions have successfully evolved into industry-grade analytics platforms, demonstrating the value of foundational research in driving innovation.

Additionally, we discuss key challenges in blockchain data analytics, including accessibility, scalability, accuracy, and interoperability. Accessibility has significantly improved, with numerous platforms now offering real-time on-chain data analytics across multiple blockchain ecosystems, including Bitcoin, Ethereum, EVM, and non-EVM blockchains. However, scalability remains a persistent issue due to the vast size of raw blockchain data, often requiring users to pay for access to insights. Academic research can address this challenge by developing efficient bootstrapping methods that maintain high accuracy while reducing data processing overhead. 
Another critical challenge is accuracy, as many industry analytics tools generate insights without clear mechanisms for data verification. Wallet labeling, in particular, presents difficulties, with some platforms relying on AI-based heuristics to classify wallet ownership and transaction behaviors. Future academic research should focus on improving these methodologies to enhance the precision and reliability of blockchain analytics. 
Lastly, interoperability remains a major challenge, particularly in cross-chain data aggregation, due to the lack of standardized protocols for integrating data across heterogeneous blockchain networks. Addressing these challenges will be essential for advancing blockchain analytics and ensuring the reliability of blockchain-based insights.

\bibliographystyle{plainurl}
\bibliography{blockchain} 

\end{document}